\documentclass[a4paper]{article}

\usepackage{INTERSPEECH2021}
\usepackage{url,subfigure,multirow,array,cite}
\usepackage{hyperref} 

\title{Enriching Source Style Transfer in Recognition-Synthesis based Non-Parallel Voice Conversion}
\name{Zhichao Wang$^1$, Xinyong Zhou$^1$, Fengyu Yang$^1$, Tao Li$^1$, Hongqiang Du$^1$, Lei Xie$^1$$^*$, \\Wendong Gan$^2$, Haitao Chen$^2$, Hai Li$^2$}

\address{
  $^1$Audio, Speech and Language Processing Group (ASLP@NPU), School of Computer Science,
  Northwestern Polytechnical University, Xi’an, China\\
  $^2$iQIYI Inc, China}
\email{zcwang\_aslp@mail.nwpu.edu.cn,\{xyzhou,fyyang,taoli,hqdu\}@nwpu-aslp.org,\\lxie@nwpu.edu.cn,\{wendonggan,haitaochen,haili\}@qiyi.com}

\begin{document}

\maketitle

\begin{abstract}
 Current voice conversion (VC) methods can successfully convert timbre of the audio. As modeling source audio's prosody effectively is a challenging task, there are still limitations of transferring source style to the converted speech. This study proposes a source style transfer method based on recognition-synthesis framework. Previously in speech generation task, prosody can be modeled explicitly with prosodic features or implicitly with a latent prosody extractor. In this paper, taking advantages of both, we model the prosody in a hybrid manner, which effectively combines explicit and implicit methods in a proposed prosody module. Specifically, prosodic features are used to explicit model prosody, while VAE and reference encoder are used to implicitly model prosody, which take Mel spectrum and bottleneck feature as input respectively. Furthermore, adversarial training is introduced to remove speaker-related information from the VAE outputs, avoiding leaking source speaker information while transferring style. Finally, we use a modified self-attention based encoder to extract sentential context from bottleneck features, which also implicitly aggregates the prosodic aspects of source speech from the layered representations. Experiments show that our approach is superior to the baseline and a competitive system in terms of style transfer; meanwhile, the speech quality and speaker similarity are well maintained.
\end{abstract}

\renewcommand{\thefootnote}{\fnsymbol{footnote}}
\footnotetext{*Corresponding author.}

\section{Introduction}

\begin{figure*}[h]
  \centering
  \includegraphics[width=0.9\linewidth]{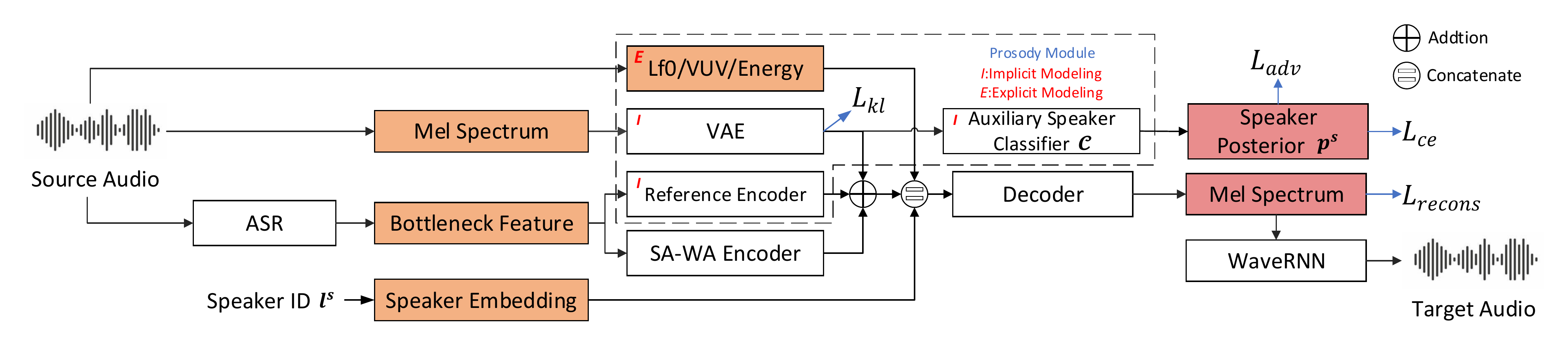}
  \caption{Overview of the components of the proposed conversion model.}
  \label{fig:model_framework}
\vspace{-8pt}
\end{figure*}

Voice conversion (VC) aims to modify speech from a source speaker to sound like that of a target speaker without changing the linguistic content. Various voice conversion approaches have been proposed before, including Gaussian mixture models (GMM)~\cite{Stylianou1998,Toda2007}, frequency warping~\cite{Erro2010,Godoy2012} and exemplar methods~\cite{Takashima2012,ZWU2014}. These methods usually require parallel data from the source and target speakers, which is expensive to collect. Therefore, many works~\cite{Hsu2016Voicevae,Sun2016PhoneticPF} propose to use non-parallel data for voice conversion. Benefiting from the deep learning's strong ability of extracting rich features and non-linear regression, non-parallel approaches based on deep learning have become the mainstream. Among them, phonetic posteriorgram (PPG)~\cite{Sun2016PhoneticPF,Tian2019ASW,Li2020PPGbasedSVC} based approaches have received much attention. The framework generally consists of a \textit{recognition} stage and a \textit{synthesis} stage, where PPG or neural network bottleneck features (BN) serve as the 'information bridge'. For the recognition stage, an ASR acoustic model is used to extract speaker-independent PPG or BN features, which represents the linguistic content. The synthesis stage is a conversion model, which maps the extracted features by ASR to the acoustic features of the target speaker. Although sequence-to-sequence framework has been recently proposed to integrate the recognition and synthesis stages into a unified encoder-decoder neural architecture that can be trained in an end-to-end manner~\cite{qian2019autovc,chou2019oneshot,Zhang2020NonParallelSV}, the classical recognition-synthesis framework is still the widely-adopted non-parallel voice conversion framework in real applications because it is \textit{flexible} -- the two stages can be separately trained and \textit{robust} -- the ASR acoustic model can be trained using plenty of data.

Despite recent progress, most voice conversion methods focus on the transformation of timbre to the target and the preservation of linguistic content of the source. However, the prosody of the source is also desired to be correctly transferred to the target. The prosody information at least includes emotion, pitch, duration, and loudness, which are important in some scenarios for voice conversion with expressive source speech, such as dubbing, live broadcasting and data enhancement. Therefore, transferring prosody of source speech to converted speech is an interesting task.

In speech generation tasks, including both voice conversion and text-to-speech, modeling prosody has two directions in general: prosody can be modeled explicitly with prosodic features, or implicitly with a latent prosody extractor. The explicit modeling methods extract prosodic features~\cite{Ming2016DeepBL,Ming2016ExemplarbasedSR,Raitio2020prosodyControl} from speech to represent prosody information. As prosodic features are intuitive, it is easy to understand the effect on prosody, and the prosody can be controlled independently via varying features. However, prosody is related to many factors, it is difficult to describe prosody through handcrafted acoustic features perfectly. For implicit modeling, the prosody extractor can be GST~\cite{Wang2018GST,Liu2020TransferringSS}, reference encoder~\cite{SkerryRyan2018Reference,Zhang2020VoiceCascading,Lian2021TowardsFP} and VAE~\cite{Zhang2019VAE}, which extract comprehensive prosody embedding from reference speech. However, the extracted prosody embedding is difficult to explain and control the different aspects of prosody. Additionally, the embedding may contain other prosody unrelated information, e.g. timbre, referred as timbre leakage. To solve this problem, gradient reversal layer (GRL)~\cite{Li2020PPGbasedSVC,Zhang2020VoiceCascading,Ganin2016GRL,Wang2021AccentAS,WangQing,sunsining} is introduced to remove speaker-related information from prosody.

In this paper, to obtain a non-parallel VC model with good performance and robustness in source style transfer, we choose a recognition-synthesis framework. To solve the prosody modeling problems caused by using explicit or implicit modeling alone, we design a hybrid prosody module to model prosody, which effectively combines explicit modeling and implicit modeling methods. The prosodic features are used to explicit model prosody, including logarithmic domain fundamental frequency (lf0) and the short-term average amplitude (energy), which model pitch and energy. According to the characteristic that mel spectrum is speaker-dependent and BN is speaker-independent but prosodic-dependent, we use VAE and reference encoder~\cite{SkerryRyan2018Reference,Lian2021TowardsFP,Zhang2019VAE,Zhang2020VoiceCascading} to implicitly model prosody, which take mel spectrum and BN as input respectively. Adversarial training strategy~\cite{Zhang2020NonParallelSV,Wang2021AccentAS} is also applied to wipe out the speaker-related information in the VAE output. To the best of our knowledge, we are the first to use BN to extract prosody representation. Additionally, we use the weighted aggregation self-attention encoder (SA-WA)~\cite{Yang2020ExploitingDS} to extract sentential context from BN, which also implicitly aggregates the prosodic aspects of source speech from the layered representations.  Experimental results show that with the proposed methods enriching source style transfer both explicitly and implicitly, our framework is clearly superior than the baseline systems.

The rest of the paper is organized as follows: Section 2 introduces our proposed methods. Section 3 presents the experiments and the comparison of our proposed approach against the baseline and ablation systems in terms of both subjective and objective measures. Section 5 concludes this paper.

\section{Proposed approach}
\label{sec:system_overview}
    
Our proposed source style transfer framework is shown in Fig.~\ref{fig:model_framework}. It is based on an encoder-decoder architecture consisting of three components: a prosody module, an SA-WA encoder and a decoder. The prosody module learns to extract speaker-independent prosody representations. The SA-WA encoder exploits sentential context, which is used to excavate the sentential context from different levels. The decoder takes prosody, content, and speaker embedding as input and the output is mel spectrum. Finally WaveRNN~\cite{Kalchbrenner2018wavernn} is adopted to reconstruct waveform from mel spectrum.

\subsection{Prosody module}

As shown in Fig.~\ref{fig:model_framework}, to extract more comprehensive prosody representations, we designed a prosody module using explicit and implicit hybrid modeling methods.

\textbf{Explicit modeling:} prosody includes duration, pitch, energy, and so on. To finely characterize and control the prosody, we extract acoustic features from the source audio as conditions for the model. Frame level's VC ensures the duration information's consistency. For pitch and energy, we extract lf0 and the short-term average amplitude (energy), respectively. We also extract the vuv, which indicates the voicing of the current frame. In implementation, we use World~\cite{Morise2016WORLDAV} to extract lf0. To reduce the differences between the extracted features caused by different speakers, we use the min-max normalization to normalize lf0 and energy to [0,1], which is useful to prevent performance degradation caused by the unseen style during the inference phase.

\textbf{Implicit modeling}: we use implicit modeling method consisting of two parts: VAE followed by an auxiliary speaker classifier and reference encoder. Inspired by~\cite{Li2020PPGbasedSVC,Liu2020TransferringSS,Lian2021TowardsFP,Zhang2020VoiceCascading}, we adopt VAE to extract the prosodic representation $\mathbf{z}$ from source audio's mel spectrum. We use the same VAE model's architecture and hyperparameters as \cite{Zhang2019VAE}. As $\mathbf{z}$ produced by mel spectrum is related to the speaker, which affects the timbre of the converted speech, we adopt an speaker classifier $\mathcal{C}$ to utilize adversarial training~\cite{Zhang2020NonParallelSV,Wang2021AccentAS} to remove speaker-related information. The process of adversarial training can be described as:
\vspace{-5pt}
\begin{equation}
  {p}^{s}=\mathcal{C}(\mathbf{z})
  \label{equation:classfier}
\end{equation}
\vspace{-5pt}
\begin{equation}
   L_{ce} = \text{CE}({p}^s, {l}^s)
   \label{equation:celoss}
\end{equation}
\vspace{-5pt}
\begin{equation}
  L_{adv} = ||{p}^s - {e}||_2^2
  \label{equation:advloss}
\end{equation}
where $l^s$ is the one-hot speaker label and ${p}^s$ is the prediction of speaker classifier. The auxiliary speaker classifier is trained with cross-entropy loss $L_{ce}$. To make the $\mathbf{z}$ indistinguishable to the speaker classifier, we minimize the $L_{adv}$ to force the speaker prediction $p^s$ to obey a uniform distribution $e=[1/S, \ldots, 1/S]$, ${S}$ is the number of speakers. Although adversarial training successfully removes speaker-related information in $\mathbf{z}$, part of prosody information is also inevitably removed. From ~\cite{Sun2016PhoneticPF,Wang2021AccentAS}, we can learn that BN is a speaker-independent representation, which contains prosody information as well. We consider using a reference encoder~\cite{SkerryRyan2018Reference} to extract speaker-independent prosodic representation directly from BN to enhance prosody information further. The outputs of VAE and reference encoder are concatenated to form the prosody embedding of our implicit modeling method.

\subsection{SA-WA encoder}

\begin{figure}[h]
  \vspace{-10pt}
  \centering
  \includegraphics[width=0.75\linewidth]{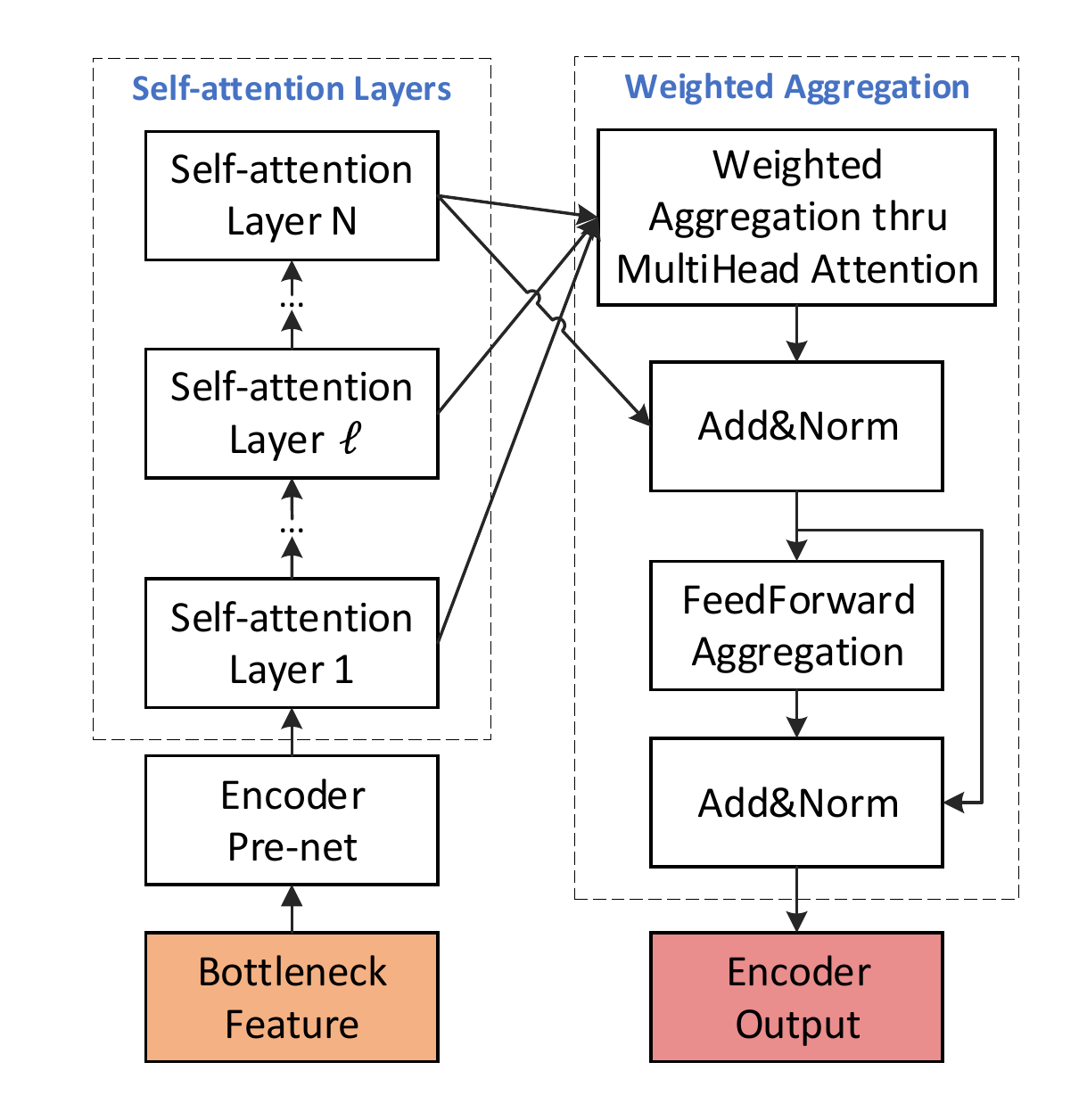}
  \vspace{-5pt}
  \caption{The network architecture of the SA-WA encoder}
  \label{fig:san_encoder}
  \vspace{-10pt}
\end{figure}

As shown in Fig.~\ref{fig:san_encoder}, we adopt an self-attention weighted aggregation (SA-WA) encoder~\cite{Yang2020ExploitingDS} to learn a comprehensive sentence representation, which is aggregated by the prosodic-related sentential context information from different self-attention layers. There are three parts in the SA-WA encoder: encoder prenet, self-attention module and weighted aggregation. The architecture of encoder prenet is the same as the settings in Tacotron~\cite{Wang2017TacotronTE}. The self-attention module contains $N$ self-attention blocks and each block consists of two sub-networks: muti-head attention and a feed forward network. Layer normalization and residual connection are also used in each sub-network. For the $n^{th}$ self-attention block, the relations among previous block $f^{n-1}$, first sub-network's output $m^{n}$ and second sub-network$f^{n}$ can be described as:

\begin{equation}
  m^{n} = \text{LN}(\text{MultiHead}(f^{n-1})+f^{n-1}),
  \label{eq1}
\end{equation}
\vspace{-5pt}
\begin{equation}
  f^{n} = \text{LN}(\text{FFN}(m^{n})+m^{n})
  \label{eq2}
\end{equation}
where MultiHead(·), FFN(·) and LN(·) are multi-head attention, feed forward network and layer normalization respectively. The weighted aggregation module consists of two sub-networks, similar to the self-attention block. Weighted aggregation can be formulated as follows:

\vspace{-8pt}
\begin{equation}
  g^{n} = \text{MeanPool}(\text{Conv1d}(f^{n})),
  \label{eq3}
\end{equation}
\vspace{-8pt}
\begin{equation}
  c = \text{LN}(\text{Multihead}(g^{1},...,g^{N})+g^{N}),
  \label{eq4}
\end{equation}
\vspace{-8pt}
\begin{equation}
  g = \text{LN}(\text{FFN}(c)+c),
  \label{eq5}
\end{equation}
where Conv1d means 1d-convolution, MeanPool represents mean pooling, the modified $c$ offers a more precise control of aggregation for each $g^{n}$ , and $g$ is the final output of encoder. 

\subsection{Loss function}
The whole conversion model is jointly optimized. We regard the auxiliary
speaker classifier as discriminator $D$, and the remaining parts as generator $G$, as shown in Fig.~\ref{fig:model_framework}. The overall loss function can be described as follows:

\begin{equation}
   \left\{ \begin{array}{l}
    Loss_G=L_{recons}+\beta L_{adv}+\gamma L_{kl}\\
    \\
    Loss_D=L_{ce}
      \end{array} \right.
  \label{equation:lossall}
\end{equation}
where $L_{recons}$ denotes the mel spectrum reconstruction loss, $\beta$ is the weight for the adversarial loss, and $\gamma$ is the variable weight for Kullback-Leibler (KL) loss to prevent VAE collapse. $G$ and $D$ are trained alternately. In general implementation, the discriminator D is updated several times for each update of the generator G. It is crucial to ensure that the classifier has a certain classification ability in the adversarial training process. To improve the model's generalization ability and the speaker similarity of converted speech, the whole network is first pre-trained on a multi-speaker dataset, and then is adapted with data from the target speaker. Note that only the decoder is optimized during the adaptation process.

\section{Experiments}

\subsection{Dataset and Experimental Setup}


A multi-speaker corpus containing 42 hours of speech from 59 speakers is used to pre-train the conversion model. A female speaker database \textit{db1}~\cite{db1} is utilized to fine-tune the conversion model. The languge is Mandarin Chinese. For conversion test, 240 utterances are used to evaluate, which contains different styles, including reading, emotion, news, novels, live broadcast, and dubbing. We use 80-dim mel spectrum as features computed in 50ms frame length and 12.5ms frame shift. Our SI-ASR acoustic system is a robust TDNN-F model trained with 30k hours of Mandarin speech data. We use the 256-dim bottleneck features as the linguistic representation, which is extracted from the last fully-connected layer before softmax. WaveRNN~\cite{Kalchbrenner2018wavernn} is adopted to reconstruct waveform from mel spectrum. The vocoder is trained using db1 speech data. 

To validate the performance of our proposed method, we implement the following systems:

\begin{itemize}
    \item \textbf{CS:} we implement a comparison system in \cite{Lian2021TowardsFP}.
	\item \textbf{BL:} baseline system follows Tacotron~\cite{Wang2017TacotronTE} framework consisting of CBHG encoder and auto-regressive decoder~\cite{Shen2018NaturalTS}. Moreover, we concatenate the lf0, vuv and energy with encoder output.
	\item \textbf{P1:} adopt VAE model with auxiliary speaker classifier $\mathcal{C}$ to extract prosody representation based on BL system. 
	\item \textbf{P2:} use SA-WA encoder instead of CBHG encoder based on P1.
	\item \textbf{P3:} the final system we proposed with SA-WA encoder and prosody module both as described in Section~\ref{sec:system_overview}.
\end{itemize}

In our implementation, for SA-WA encoder, we use six self-attention blocks, each of which contains two heads of multi-head attention and two linear transformations with 512 and 256 hidden units and the settings of other parts remain the same as \cite{Yang2020ExploitingDS}. The architecture and hyperparameters of VAE and reference encoder keep the origin configuration~\cite{Zhang2019VAE,SkerryRyan2018Reference}. The speaker classifier consists of 3 fully connected layers. The decoder adopts an auto-regressive module~\cite{Shen2018NaturalTS} consisted of prenet, decoder RNN and postnet. In pre-train and fine-tuning stages, the conversion model is trained for 250k steps using batch size of 16, respectively. We use Adam optimizer to optimize our model with learning rate decay, which starts from 0.0002 and decays every 25k steps in decay rate 0.5. The generator $G$ and discriminator $D$ are set to alternate every 5 steps of training, and $\beta$ is set to 0.1.

\subsection{Listening test}

We conduct the following listening tests: two mean opinion score (MOS) tests to assess audio quality and speaker similarity respectively, and an AB test to evaluate the performance of source style transfer. We randomly select 15 utterances from each system. Twelve listeners participated in the tests. We highly recommend the readers to listen to our samples\footnote{Samples can be found in \href{https://kerwinchao.github.io/SourceStyleTransfer.github.io/}{\url{https://kerwinchao.github.io/SourceStyleTransfer.github.io/}}}.

\textbf{Quality and similarity}. For the two MOS tests, listeners are asked to rate the quality and speaker similarity of the converted speech on a 5-point scale. Additionally, the speech from target speaker (TA) is also evaluated with MOS tests. The results of MOS tests are shown in Table~\ref{tab:mos}. It is observed that the speech quality of the five systems is not significantly different. For speaker similarity, BL gets a higher score, and CS achieves a slightly poor result. Since the transfer of source style may affect the timbre and there is no ground truth, these two factors may affect the listeners' judgment. These results show that our proposed method has stability in audio quality and timbre.
\begin{table}[h]
\vspace{-5pt}
  \caption{MOS results with 95\% confidence interval.}
  \vspace{-5pt}
  \label{tab:mos}
  \centering 
  \scriptsize
  \renewcommand\arraystretch{1.5}
  \begin{tabular}{m{0.2cm}<{\centering}|m{2.4cm}|m{1.5cm}<{\centering}m{2cm}<{\centering}}
  \hline
  \textbf{ID} &\textbf{Model}                 &\textbf{Quality} &\textbf{Speaker Similarity}       \\ \hline
    TA &Target speaker speech                        & 4.41$\pm$0.051                               & 4.25$\pm$0.042                                       \\ \hline
CS &Comparison system\cite{Lian2021TowardsFP}                       &3.51$\pm$0.066                               &3.51$\pm$0.068                                       \\ \hline
  BL &Baseline                        & 3.56$\pm$0.068                               & 3.65$\pm$0.068                                       \\ \hline
  P1 &\ + VAE with classifier $\mathcal{C}$                      & 3.53$\pm$0.079                               & 3.61$\pm$0.074                                       \\ \hline
  P2 &\ \ \ + SA-WA encoder & 3.53$\pm$0.087                      & 3.58$\pm$0.088  \\ \hline
  P3 &\ \ \ \ + prosody module & 3.54$\pm$0.081                      & 3.59$\pm$0.083   \\ \hline
  \end{tabular}
  \end{table}
 
\begin{figure}[h]
    \centering
    \includegraphics[width=1.0\linewidth]{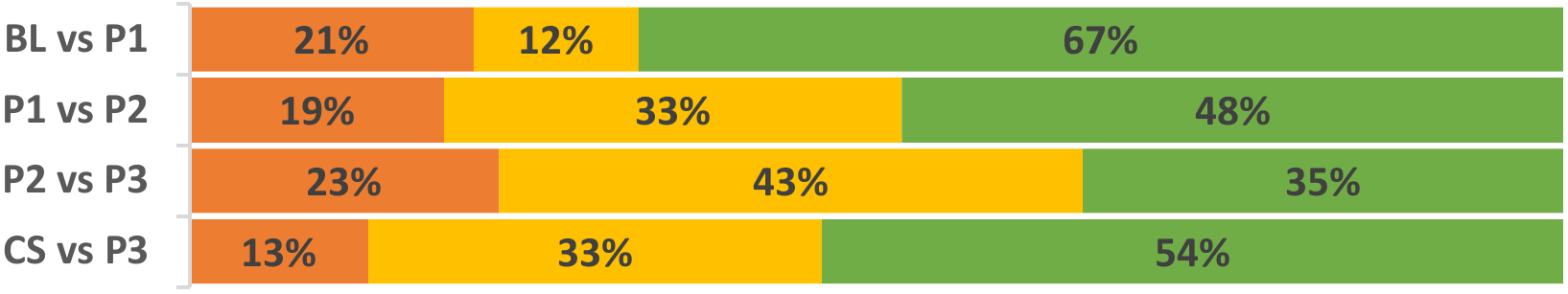}
    \vspace{-15pt}
  \caption{Source style transfer preference test results (A:NP:B).}
  \label{fig:ABTEST}
\end{figure}

\textbf{Source style transfer preference}. In the AB test, listeners are asked to choose from paired samples with better style transfer performance. The results are shown in Fig.~\ref{fig:ABTEST}. Comparing with CS, P3 gets a clearly better result. And with the addition of new components and methods, the performance of the source style transfer has also improved. The results validate that both SA-WA encoder and prosody module are beneficial to the perceived performance. Our proposed P3 achieves the best performance.
  \vspace{-10pt}
\vspace{-15pt}

\subsection{Objective Measures}

\textbf{Prosody correlation}. To further verify the statistical significance of the expressiveness of each system, we extracted features related to the prosody: frame-level energy and lf0. First, we used 240 samples to calculate the Pearson correlation coefficients between all systems. The higher the Pearson correlation coefficient of the model, the higher the accuracy of the predicted prosodic attributes. As shown in Table~\ref{tab:pearson}, P3 gets the highest scores from the perspective of energy and lf0.

\begin{figure}[h]
  \centering
  \subfigure[Energy]{
  \begin{minipage}[t]{0.5\linewidth}
  \centering
  \includegraphics[width=1.55in]{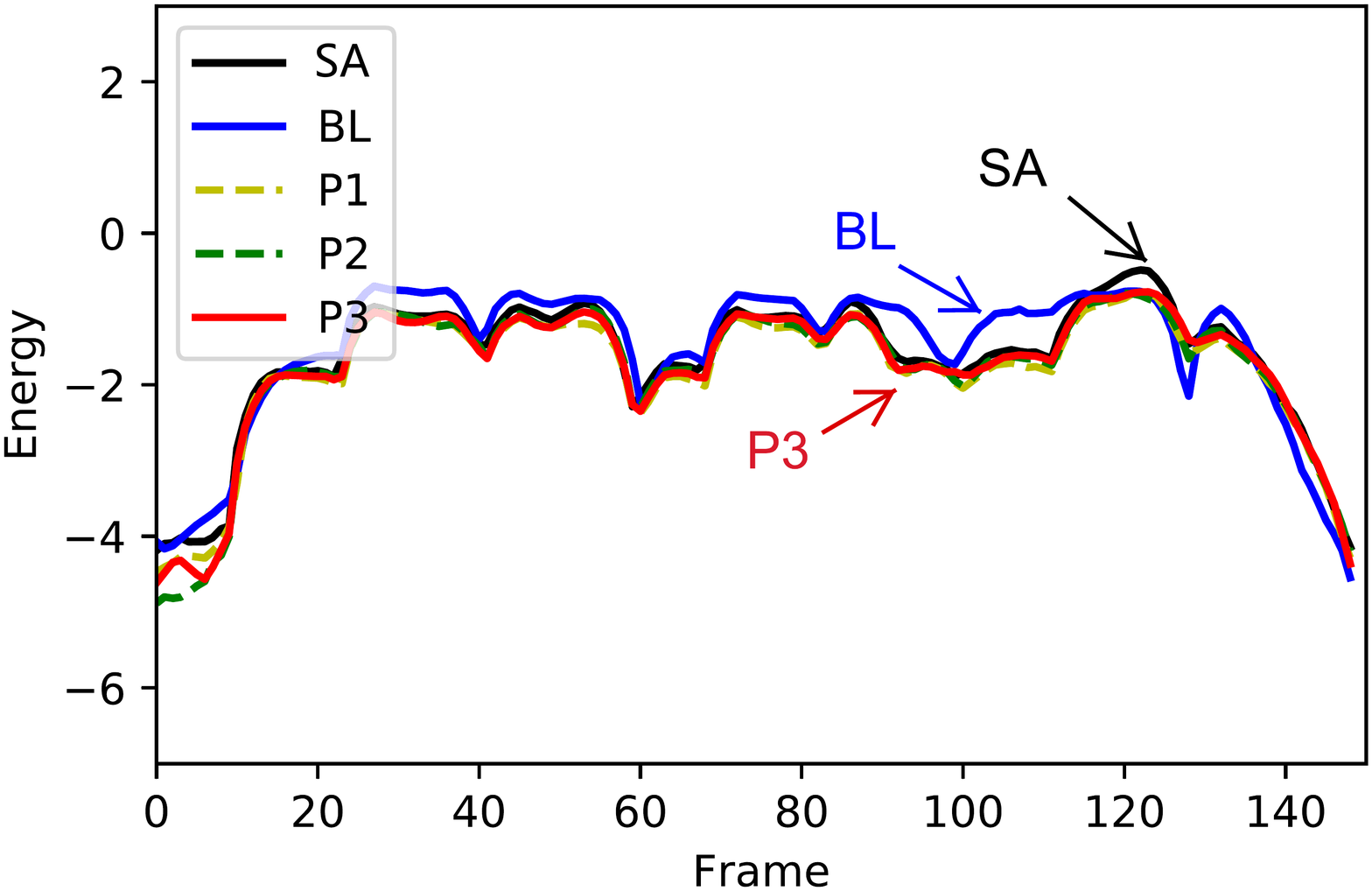}
  \end{minipage}%
  }%
  \subfigure[F0]{
  \begin{minipage}[t]{0.5\linewidth}
  \centering
  \includegraphics[width=1.55in]{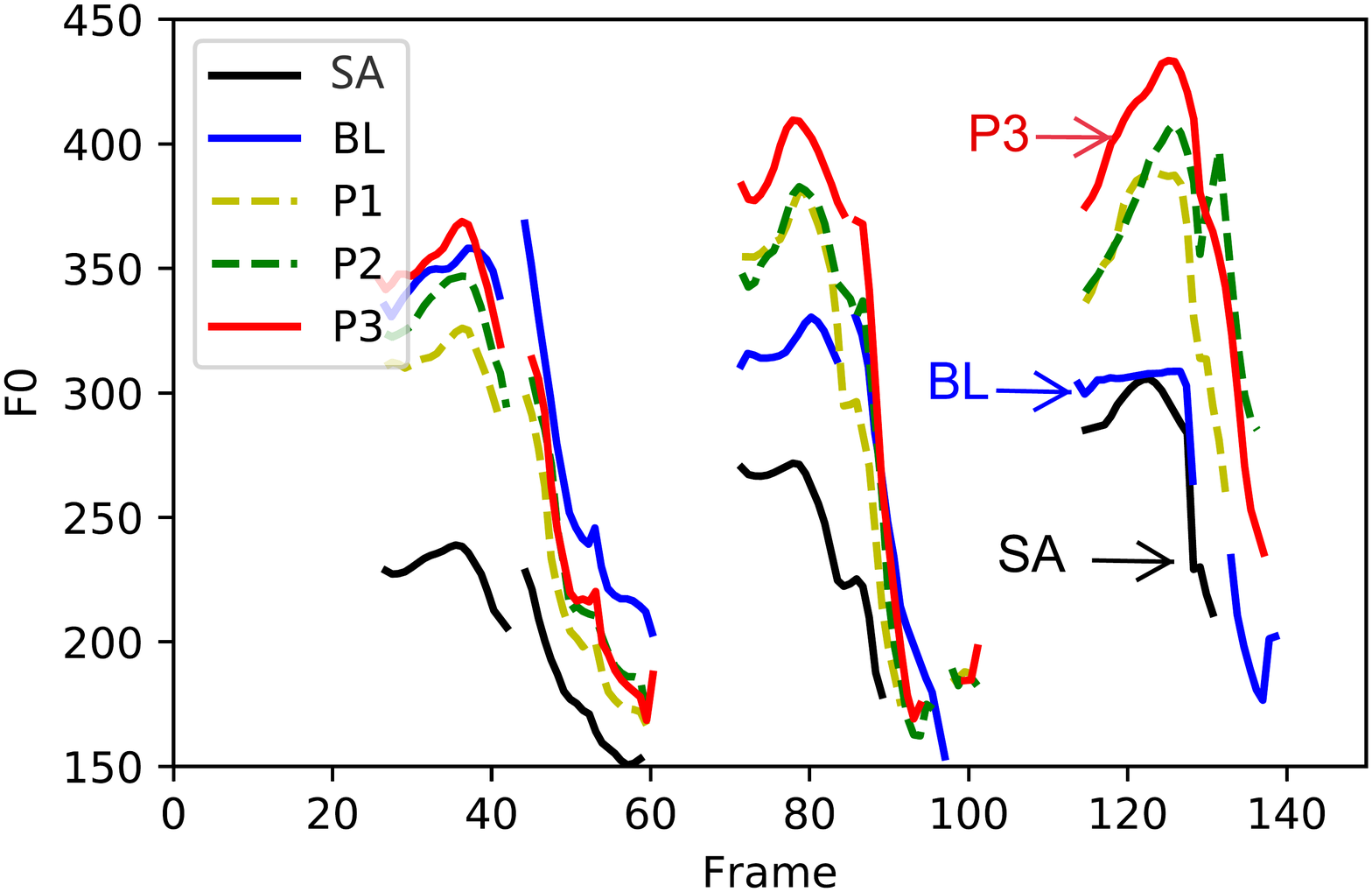}
  \end{minipage}%
  }%
  \vspace{-5pt}
\caption{Energy and f0 values of a converted speech generated by different systems. Please focus on the trend of the curve.}
\label{fig:diagram}
\vspace{-15pt}
\end{figure}

Simultaneously, as shown in Fig.~\ref{fig:diagram}, we have drawn f0 and energy trajectory diagrams of converted speeches and source audio (SA). Fig.~\ref{fig:diagram} (a) shows that all other systems achieve similar energy curves as SA except for BL. Fig.~\ref{fig:diagram}(b) shows a significantly different trend between BL and SA in the curve trend while the other systems keep the same. Among them, P3 achieves a more extensive pitch range and more potent expressiveness. These results above indicate that our proposed system P3 models the source style better.
\begin{table}[h] 
\centering
\vspace{-5pt}
\caption{Correlation in relative energy and lf0.}
\begin{tabular}{m{1cm}<{\centering}|m{1cm}<{\centering}m{1cm}<{\centering}m{1cm}<{\centering}m{1cm}}
\hline
\label{tab:pearson}
\vspace{-5pt}
\textbf{} & BL    & P1    & P2    & P3             \\ \hline
Energy    & 0.951 & 0.970 & 0.970 & \textbf{0.973} \\ \hline
Lf0       & 0.665 & 0.777 & 0.786 & \textbf{0.793} \\ \hline
\end{tabular}
\end{table}
\vspace{-5pt}

\begin{figure}[h]
  \centering
  \subfigure[Energy]{
  \begin{minipage}[t]{0.5\linewidth}
  \centering
  \includegraphics[width=1.55in]{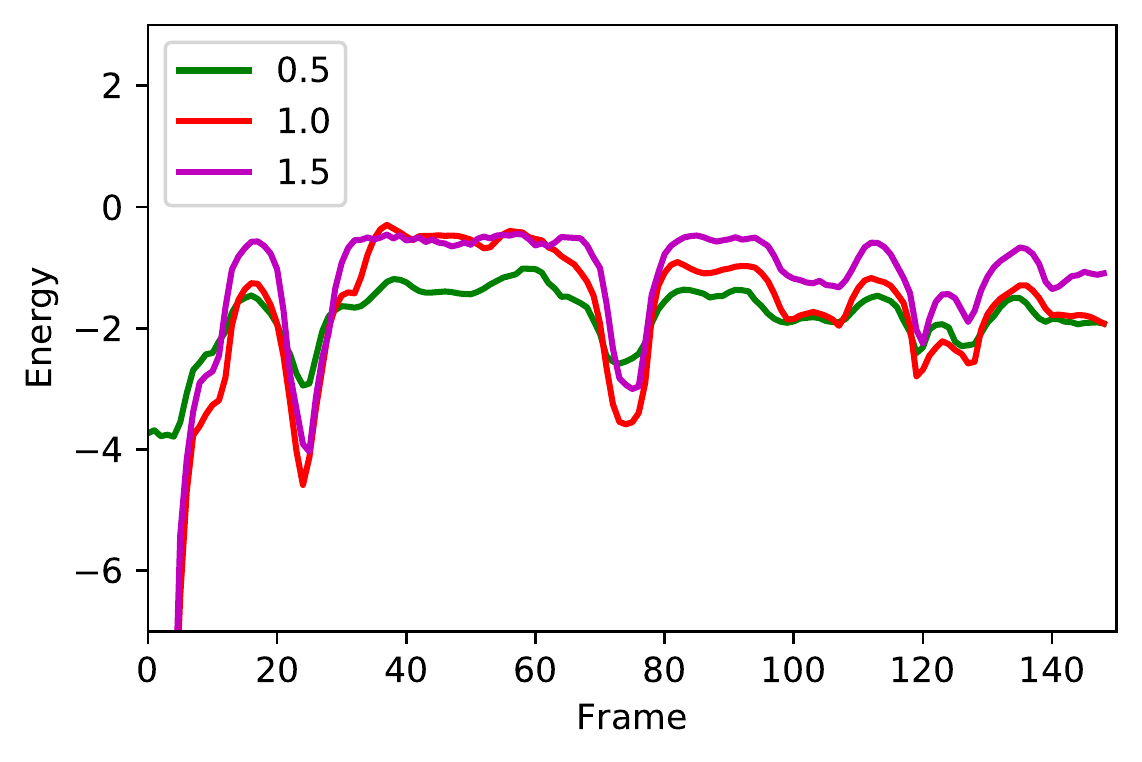}
  \end{minipage}%
  }%
  \subfigure[F0]{
  \begin{minipage}[t]{0.5\linewidth}
  \centering
  \includegraphics[width=1.55in]{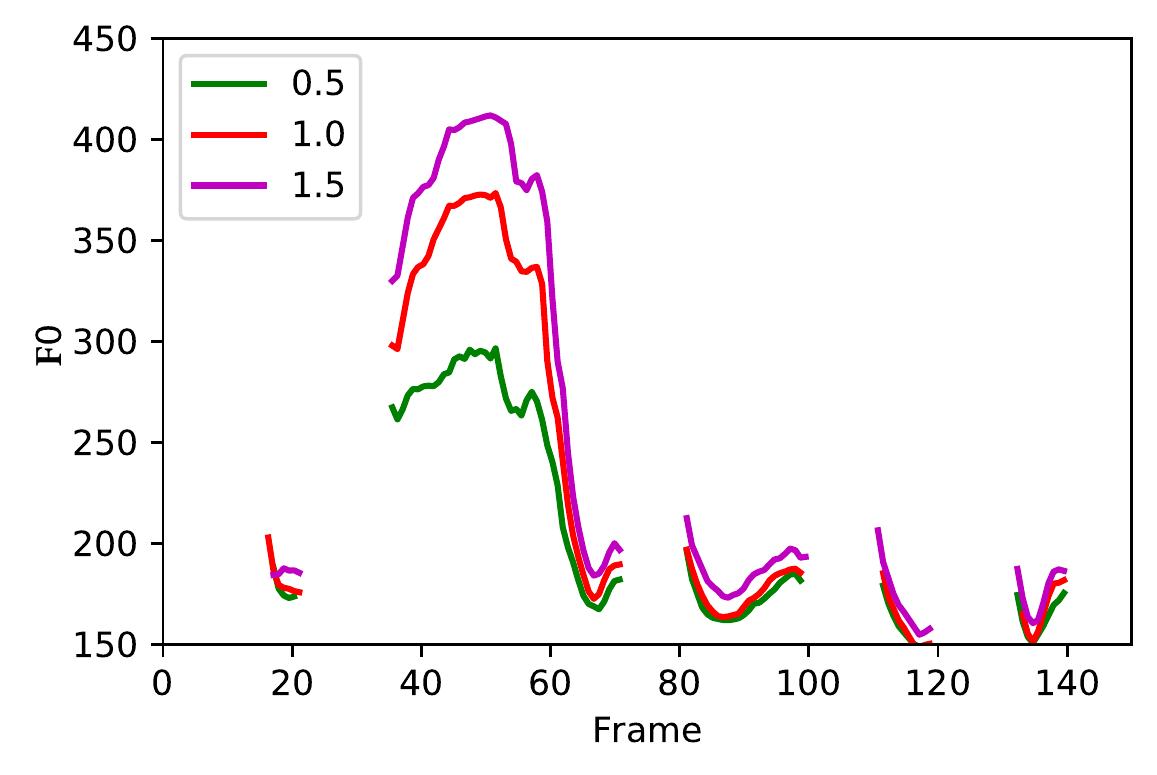}
  \end{minipage}%
  }%
  \vspace{-5pt}
\caption{Intuitive prosody control of energy and f0.}
\label{fig:scale}
\vspace{-5pt}
\end{figure}

\textbf{Prosody control}. To verify whether lf0 and energy work in the model, we assume that converted speech can be easily controlled through intuitive control of the prosody characteristics. To test this hypothesis, the lf0 and energy of one sentence from test data are multiplied by the coefficients of 0.5, 1 and 1.5, respectively. And we extracted the converted speeches' f0 and energy for visualization. The results are shown in Fig.~\ref{fig:scale}. It can be seen from the figure that as the coefficient increases, the values and fluctuations of the energy and f0 curves become larger. Meanwhile, there is no side effect in the converted speech's timber. This result indicates that lf0 and energy are well-used and can intuitively control the related prosody in our model. 

\begin{figure}[h]
  \centering
  \vspace{-7pt}
  \subfigure[w/o ADV]{
  \begin{minipage}[t]{0.5\linewidth}
  \centering
  \includegraphics[width=1.55in]{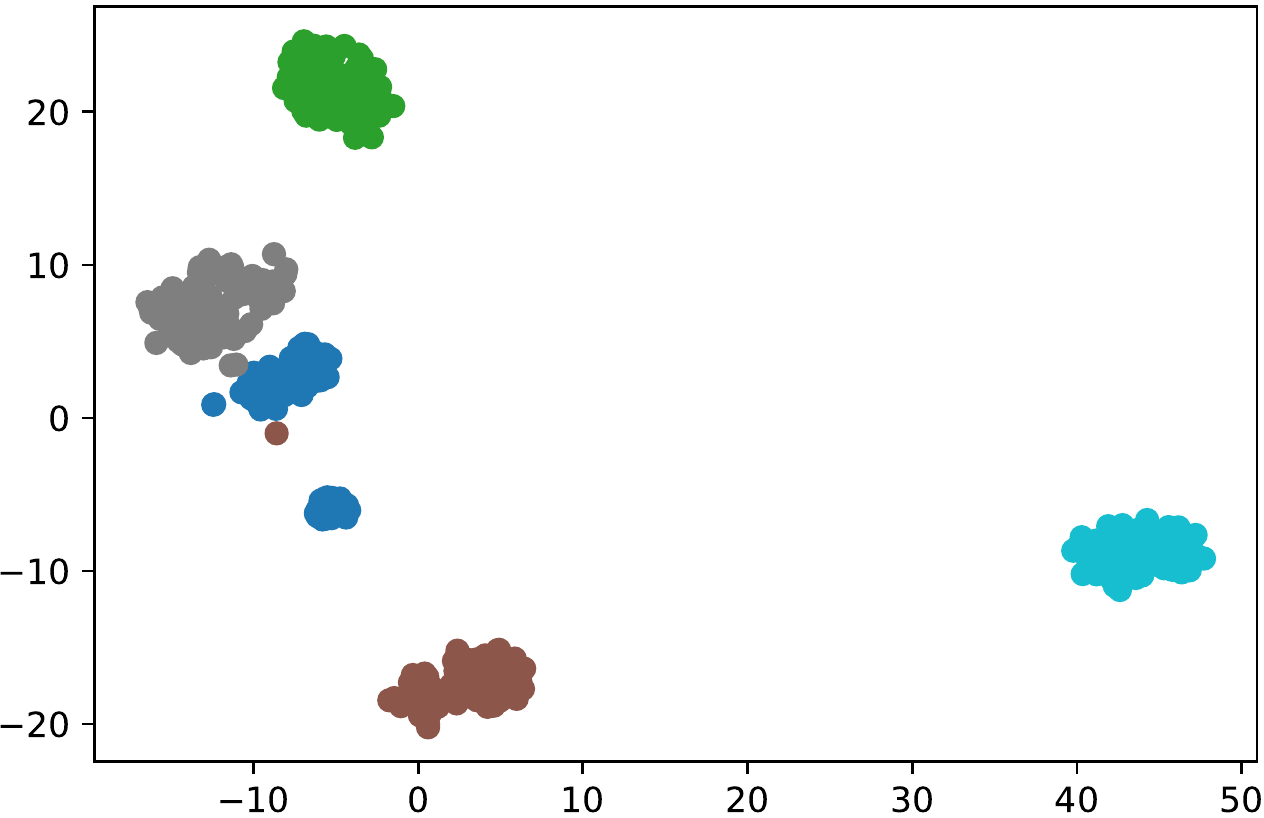}
  \end{minipage}%
  }%
  \subfigure[w/ ADV]{
  \begin{minipage}[t]{0.5\linewidth}
  \centering
  \includegraphics[width=1.55in]{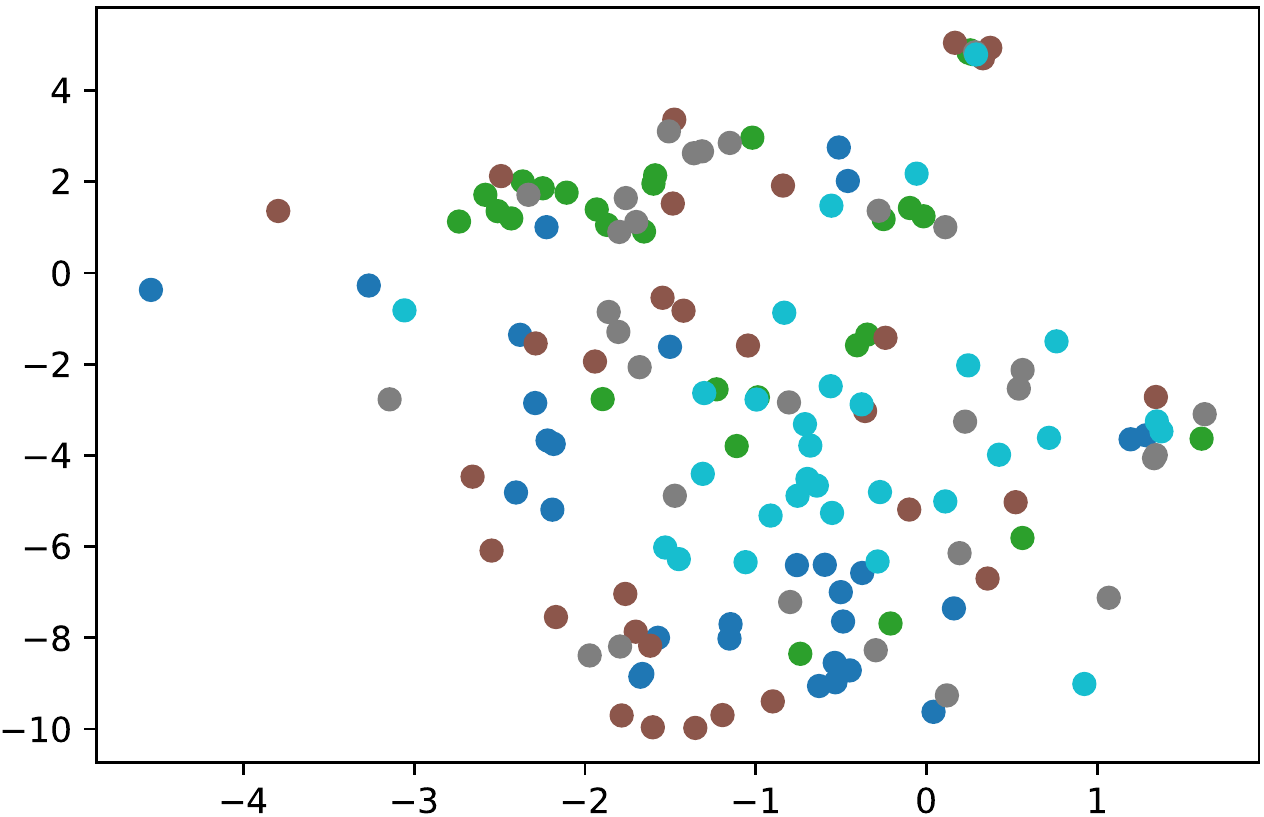}
  \end{minipage}%
  }%
  \vspace{-5pt}
  \caption{VAE output visualization using t-SNE. Each color represents a speaker.}
\label{fig:t-sne}
\vspace{-10pt}
\end{figure}

\textbf{Visualization of VAE Output}. To demonstrate the ability to remove speaker-related information using adversarial training strategy (ADV), we visualize the VAE outputs by t-SNE~\cite{VanDerMaaten2008} with and without using ADV in Fig.~\ref{fig:t-sne}. Thirty utterances from 5 source speakers are selected. The VAE outputs are projected to 2D by t-SNE~\cite{VanDerMaaten2008}. Each color represents a speaker. With the help of adversarial training, the outputs from different speakers are uniformly distributed in Fig.~\ref{fig:t-sne} (b). This indicates that adversarial training effectively helps VAE to generate speaker-independent outputs.


\section{Conclusions}

In this study, we propose a hybrid modeling approach for source style transfer in voice conversion based on recognition-synthesis framework, which can transfer the style of source speech to the converted speech. This task is challenging as the speaker’s timbre and prosodic style are coupled with each other, and it is difficult to get enough rich and speaker-independent prosody information. Specifically, we first model the prosody in a hybrid manner, which combines implicit and explicit modeling methods. Furthermore, adversarial training is introduced to remove speaker-related information from the extracted prosody. Finally, we use SA-WA encoder to extract comprehensive linguistic content representation. Experimental results show that our proposed system has better abilities in modeling prosody-related acoustic features and controlling synthesized speech prosody. 

\bibliographystyle{IEEEtran}

\bibliography{mybib}

\end{document}